\documentclass[prb, twocolumn]{revtex4-1}
\usepackage{epsf,epsfig}
\usepackage{amsmath}
\usepackage{natbib}
\usepackage[percent]{overpic}
\usepackage{float}

\usepackage[export]{adjustbox}
\usepackage[toc,page]{appendix}
\usepackage{graphicx,placeins, tikz}
\usepackage[caption=false]{subfig}

\usepackage{braket} %braket
\usepackage[version=3]{mhchem}

\usepackage{bm}% bold math

\newcommand {\apgt} {\ {\raise-.5ex\hbox{$\buildrel>\over\sim$}}\ }
\newcommand {\aplt} {\ {\raise-.5ex\hbox{$\buildrel<\over\sim$}}\ }
\newcommand{\lessim}{\aplt}
\newcommand{\gssim}{\apgt}

\newcommand{\disp}[1]{Eq.~(\ref{#1})}
\newcommand{\refdisp}[1]{Ref.~(\onlinecite{#1})}
\newcommand{\figdisp}[1]{Fig.~(\ref{#1})}

\usepackage{color}
\usepackage{bm}% bold math
\usepackage{alltt,dsfont}
\usepackage{appendix}
\usepackage{amsmath,amssymb}
\usepackage{hyperref}
\hypersetup{
    colorlinks,
    citecolor=red,
    filecolor=cyan,
    linkcolor=blue,
    urlcolor=magenta
}

\usepackage{atveryend}
\makeatletter
\let\origcitation\citation
\AtEndDocument{\def\mycites{}%
  \def\citation#1{\g@addto@macro\mycites{#1^^J}\origcitation{#1}}}
\AtVeryEndDocument{\newwrite\citeout\immediate\openout\citeout=\jobname.cit
  \immediate\write\citeout{\mycites}\immediate\closeout\citeout}
\makeatother

\newcommand*{\shifttext}[2]{%
  \settowidth{\@tempdima}{#2}%
  \makebox[\@tempdima]{\hspace*{#1}#2}%
}

\newcommand{\beq}{\begin{equation}}
\newcommand{\eeq}{\end{equation}}
\newcommand{\barray}{\begin{eqnarray}}
\newcommand{\earray}{\end{eqnarray}}

\newcommand{\nn}{\nonumber}

\begin{document}
\title{Kondo-Ising and Tight-Binding Models for $\ce{TmB_4}$ }
\author{John Shin, Zack Schlesinger, B Sriram Shastry}
\affiliation{Physics Department, University of California,  Santa Cruz, Ca 95064 }
\date{\today}
\begin{abstract}
In $\ce{TmB_4}$, localized electrons with a large magnetic moment interact with metallic electrons in boron-derived bands. We examine the nature of $\ce{TmB_4}$ using full-relativistic ab-initio density functional theory calculations, approximate tight-binding Hamiltonian results, and the development of an effective Kondo-Ising   model for this system. Features of the Fermi surface relating to the anisotropic conduction of charge  are discussed.
 The observed magnetic moment $\sim 6 \, \mu_B$ is argued to require a  subtle   crystal field effect in metallic systems,  involving  a flipped sign of the effective charges surrounding a Tm ion.
The  role of on-site   quantum dynamics  in the resulting Kondo-Ising type ``impurity''  model are highlighted. From this model,  elimination of the conduction electrons will lead to spin-spin  (RKKY-type)  interaction of Ising character required  to  understand the observed fractional magnetization plateaus in $\ce{TmB_4}$.
\end{abstract}
\maketitle

\section{Introduction}

$\ce{TmB_4}$ is a metallic frustrated magnet, with an Ising-like anti-ferromagnetic ground state\cite{Siemensmeyer}. Experiments have shown a variety of rich phenomena--along with the fractional magnetization plateaus at low temperature (occurring at fractions of the saturation magnetization, with a stable plateau at 1/2, and fractional plateaus at 1/7, 1/8, 1/9, ... \cite{Siemensmeyer}) and its rich phase diagram, hysteretic magnetoresistance and the anomalous Hall effect have also been observed \cite{sunku}. Fractional magnetization plateaus for 2D quantum spin systems were first observed in $\ce{SrCu_2(BO_3)_2}$ \cite{SrCu} and later observed in the family of rare-earth tetraborides,\cite{Siemensmeyer} where the rare-earth ions can be mapped onto the Shastry-Sutherland lattice \cite{SS-lattice} and the boron atoms can be grouped into $\ce{B_6}$ octahedra and dimer pairs \cite{yin, lipscomb}. In contrast to the high fields necessary to reach magnetic saturation in $\ce{SrCu_2(BO_3)_2}$ \cite{SrCu}, $\ce{TmB_4}$ saturates at fields on the order of 4 T \cite{matas}, and also exhibits long-range order \cite{wierschem}. This magnetic analogue to the fractional quantum hall effect and relatively simple example of geometric frustration has spurred both theoretical and experimental interest, where the juxtaposition of 2D magnetism and 3D conduction \cite{sunku, ye2016} remains novel.

The primary focus of this paper is on the electronic and magnetic characteristics of $\ce{TmB_4}$, however, $\ce{TmB_4}$ also has interesting structural aspects and it is worthwhile to take a moment to review them. The thulium atoms lie in sheets oriented perpendicular to the tetragonal $c$-axis. Their structure can be viewed in terms of tiling of squares and triangles in a classical manner examined by Archimedes. Between these Tm sheets there are planes of boron atoms. Boron, adjacent to carbon in the periodic table, might be expected to form 6-element rings, as indeed it does in $\ce{TmB_2}$ \cite{tmb2}, however, in $\ce{TmB_4}$ boron atoms form {\it 7-atom rings} that lie in planes consisting of the 7-atom rings and squares.

The $\ce{Tm}$ is nominally trivalent and has a $\ce{4f^{12}}$ configuration. This leads to a two-hole state on the Tm with a local-moment of $M_J=\pm 6$. This local moment interacts and hybridizes with conduction electrons associated with boron bands. This paper seeks to capture the essence of that interaction and hybridization. It includes an ab-initio approach, tight-binding model calculations associated with reduced structures, and, most notably, the development of an effective Kondo Ising model for which an understanding of the symmetry of the $\ce{Tm}$ site is critical.

\section{Structure}

\begin{figure*}[t!]
\subfloat[]{%
  \includegraphics[width=0.75\columnwidth]{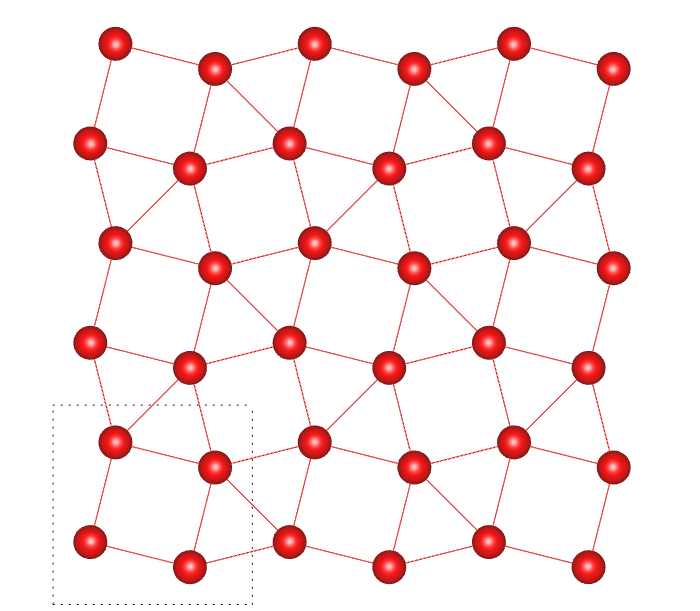}%
\label{tm_lattice}
}
\subfloat[]{%
  \includegraphics[width=0.75\columnwidth]{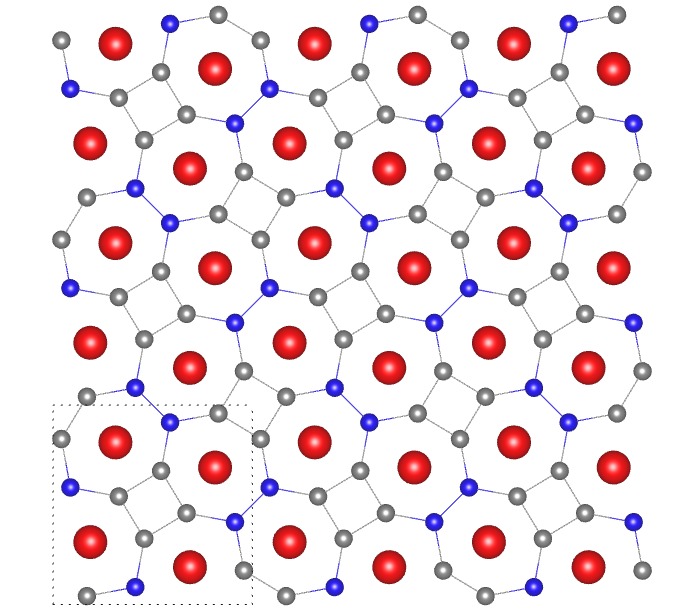}%
\label{planar_lattice}
}

\subfloat[]{%
  \includegraphics[width=0.75\columnwidth]{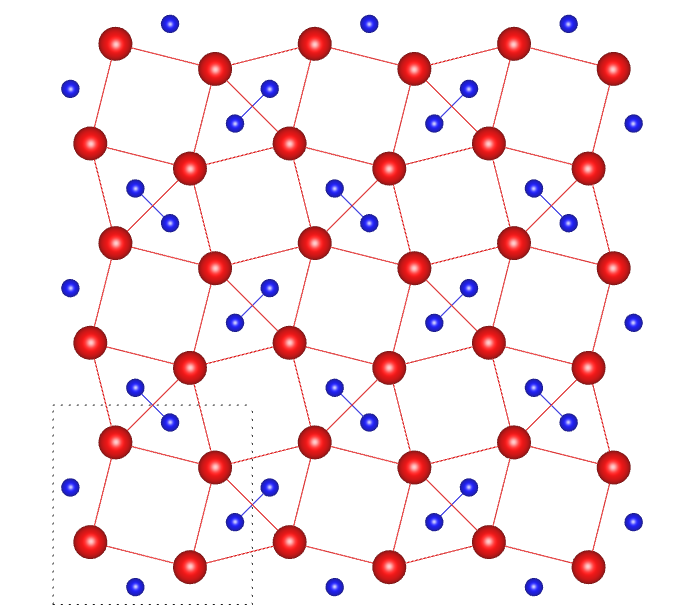}%
\label{dimer_lattice}
}
\subfloat[]{%
  \includegraphics[width=0.75\columnwidth]{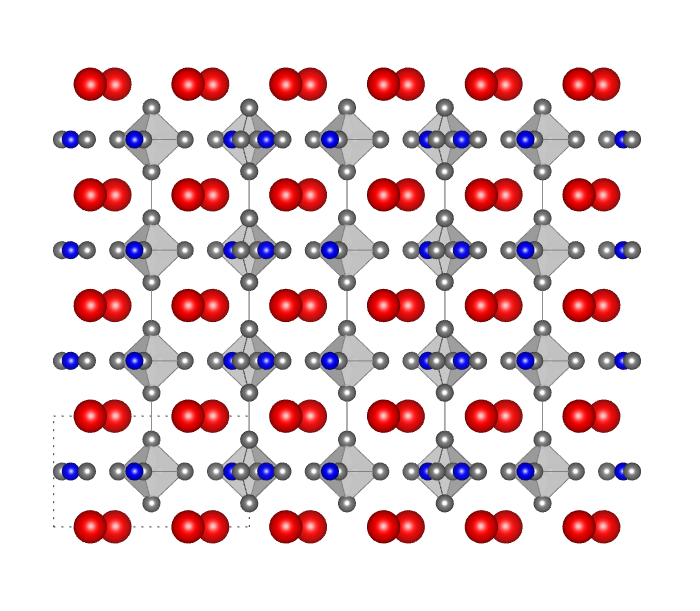}%
\label{full_lattice}
}
\caption{a) Figure 1(a) shows the structure of a single plane of $\ce{Tm}$ atoms within crystalline $\ce{TmB_4}$. Note the triangle and square tiling. b) Figure 1(b) shows the juxtaposition of a plane of $\ce{Tm}$ atoms (red) and a plane of B (blue and gray) atoms as viewed from above. The $\ce{Tm}$ plane is .20 nm above the boron atoms plane. Coupling of the Tm $\ce{f}$-orbital to the planar boron $\ce{2p_z}$ band plays an essential role in the physics and phenomenology of the system. Note the 7 atom boron rings.  c) In Figure 1(c) only the dimer boron atoms and the Tm atoms are shown. This figure shows the most simplified structure able to capture the essence of the magnetism and coupling in $\ce{TmB_4}$. d) Figure 1(d) shows the structure of $\ce{TmB_4}$ as viewed from the side (along the $a$-axis). One sees the stacking of the $\ce{Tm}$ layers in red, the blue and gray boron planes and in addition the chain formed by boron octahedra oriented along the $c$-axis. The apical boron are colored gray.}
\label{structure}
\end{figure*}

$\ce{TmB_4}$ crystallizes in a tetragonal structure (space group $P4/m b m$) \cite{lattice}, and has a mixture of 2D and 3D aspects. The Tm lattice viewed on its own consists of stacked 2D sheets, with each sheet a 2D Shastry-Sutherland lattice \cite{SS-lattice}. Each sheet has a structure that includes perfect squares and nearly equilateral triangles of $\ce{Tm}$ atoms, as shown in Fig.~(\ref{structure}\subref{tm_lattice}). Each $\ce{Tm}$ ion has 5 near $\ce{Tm}$ neighbors in the plane with angles between near neighbor bonds of 90, 59, 59, 90 and 62 degrees, where 90 is the interior angle of a square and 59 and 62 are the interior angles of an almost equilateral triangle. These $\ce{Tm}$ sheets lie in the crystalline $a$-$b$ plane and are stacked along the $c$-axis. The distance between $\ce{Tm}$ sheets is .399 nm.

Boron planes lie halfway between the $\ce{Tm}$ sheets. The structure of a boron plane involves a mixture of 7 atom rings and 4 atom squares as shown in Fig. ~(\ref{structure}\subref{planar_lattice}). There are two distinct types of boron sites in these planes. One type, shown in blue, is solely part of the boron plane. The other type, shown in light gray, is part of the boron plane and also of an octahedral chain along the $c$-axis. This second type comes in groups of 4 atoms which form a square in the plane (Fig.~(\ref{structure}\subref{planar_lattice}) and are part of an octahedron which includes apical boron atoms which are not in the plane and are thus not shown in Fig.~(\ref{structure}\subref{planar_lattice}). 

On the other hand, the pure planar boron atoms (blue) come in dimer pairs as shown in Fig.~(\ref{structure}\subref{planar_lattice}). There are no extra-planar boron above or below them, thus their $\ce{2p_z}$ orbitals are unencumbered. It is the partially occupied $\ce{2p_z}$ orbital of these pure planar boron atoms that couple most strongly to Tm $\ce{4f}$ level electrons. Most of the essential nature of the hybridization of $\ce{TmB_4}$ can be captured by studying the structures shown in Fig.~(\ref{structure}\subref{planar_lattice}) and Fig.~(\ref{structure}\subref{dimer_lattice}). In a later section we discuss tight binding results for these structures and compare those to our DFT results. One of our primary goals is to see how far one can go along the path of structural simplification while still capturing the essential nature of the interaction between the quasi-localized $\ce{f}$-state and itinerant $\ce{2p}$ states and thus the magnetic character and essential phenomenology of $\ce{TmB_4}$.
 
Fig.~(\ref{structure}\subref{full_lattice}) shows the full 3D structure of $\ce{TmB_4}$. This includes the alternating boron and Tm planes, as well as apical boron atoms (gray) which lie between the boron and Tm planes. These are located above and below squares of B atoms in the plane and form the tops and bottoms of the boron octahedra which stack along the $c$-axis creating chains which extend throughout the crystal. These chains of boron octahedra are shown to not couple strongly to the $\ce{Tm}$ orbitals and thus do not play a large role in the magnetic character of $\ce{TmB_4}$.

\begin{figure}
\includegraphics[width=.75\columnwidth]{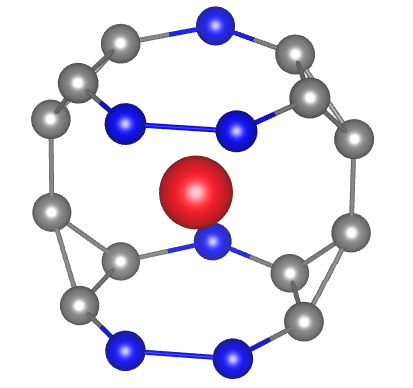}%
\caption{This shows Tm (red) and its local environment. The blue spheres indicate the dimer boron, while the gray spheres indicate the boron that participate in the formation of octahedra.}
\label{local}
\end{figure}

The local environment of $\ce{Tm}$ can be described by a model as shown in Fig.~(\ref{local}). Above and below the Tm are 7 atom boron rings, 3 of which are the previously mentioned dimer borons. Apical boron lie above and below the boron planes. There are mirror plane symmetries through the Tm plane and perpendicular to the dimer-dimer bond, and a total of 18 borons construct the local environment.

\section{Density Functional Theory Calculations}

To begin our exploration of the electronic and magnetic characteristics of $\ce{TmB_4}$, we have performed spin-polarized full relativistic density functional theory \cite{kohn} calculations \cite{xsede} with the full potential local orbital (FPLO) code (version 14.00-48) \cite{FPLO}, using the Generalized Gradient Approximation (GGA), along with an intra-atomic Hubbard U repulsion term. The PBE96 \cite{gga} and Atomic Limit functionals \cite{fullylocalizedlimit, Erik} were used for the GGA and U, respectively, with a k-mesh of $24^3$ in the Brillouin Zone. Values of $U = 8$ eV and $J = 1$ eV were used on the Tm $\ce{4f}$ orbitals. The lattice constants and atomic positions were taken from experiment \cite{lattice}. An LDA+U study of rare-earth tetraborides has previously been undertaken in \refdisp{yin}, but here we focus on the case of $\ce{TmB_4}$. We note that there are limitations of the LDA+U method in fully describing highly correlated systems \cite{eschrig}, which is exemplified by a reduced magnetic $\ce{Tm}$ moment of $M_J=3.38$ found in our meta-stable ground state compared to experiment \cite{Siemensmeyer}, but the method has been used with success in other lanthanide systems \cite{coleman, tmb2, lev, cerium}.

Many initial occupation matrix configurations \cite{occupation} were investigated, but the results shown here depict the convergence of a calculation with two initial holes starting in the high moment $\ket{7/2, \pm 7/2}$ and $\ket{7/2, \pm 5/2}$ states, with induced ferromagnetic order to supply a non-zero moment. Since the Hund's rules ground state \cite{yin} and anti-ferromagnetic order can break the crystal space group symmetry, we have also investigated reductions in symmetry with both ferromagnetic and anti-ferromagnetic order, but none converged to a metastable solution. For the correlated $\ce{4f}$ orbitals, we find a configuration of 4f12.03, and thus an inference of about 1.97 holes per $\ce{Tm}$ in high-spin $\ce{f}$-states.

\begin{figure*}[]

\subfloat{
\put(10,165){a)}
\includegraphics[width=1\columnwidth]{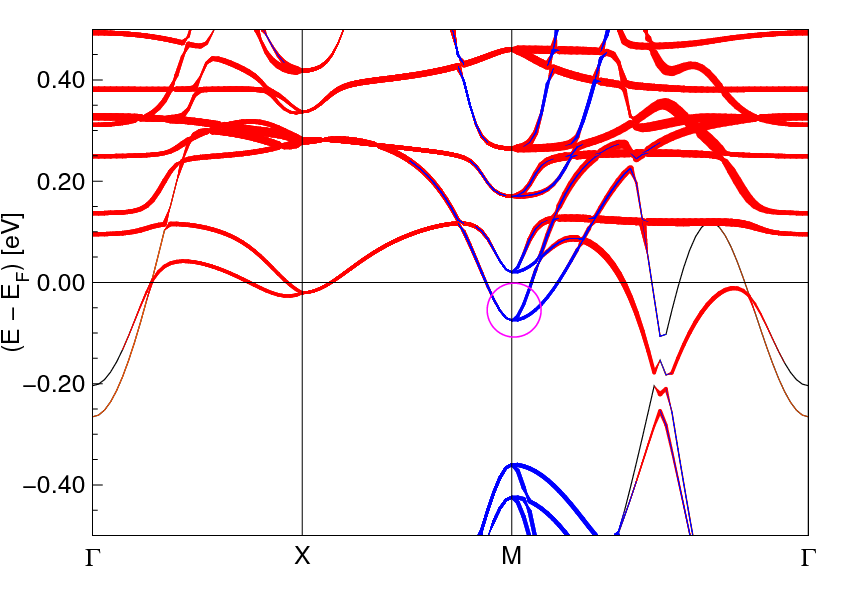}\llap{\raisebox{.5cm}{\includegraphics[height=2.5cm]{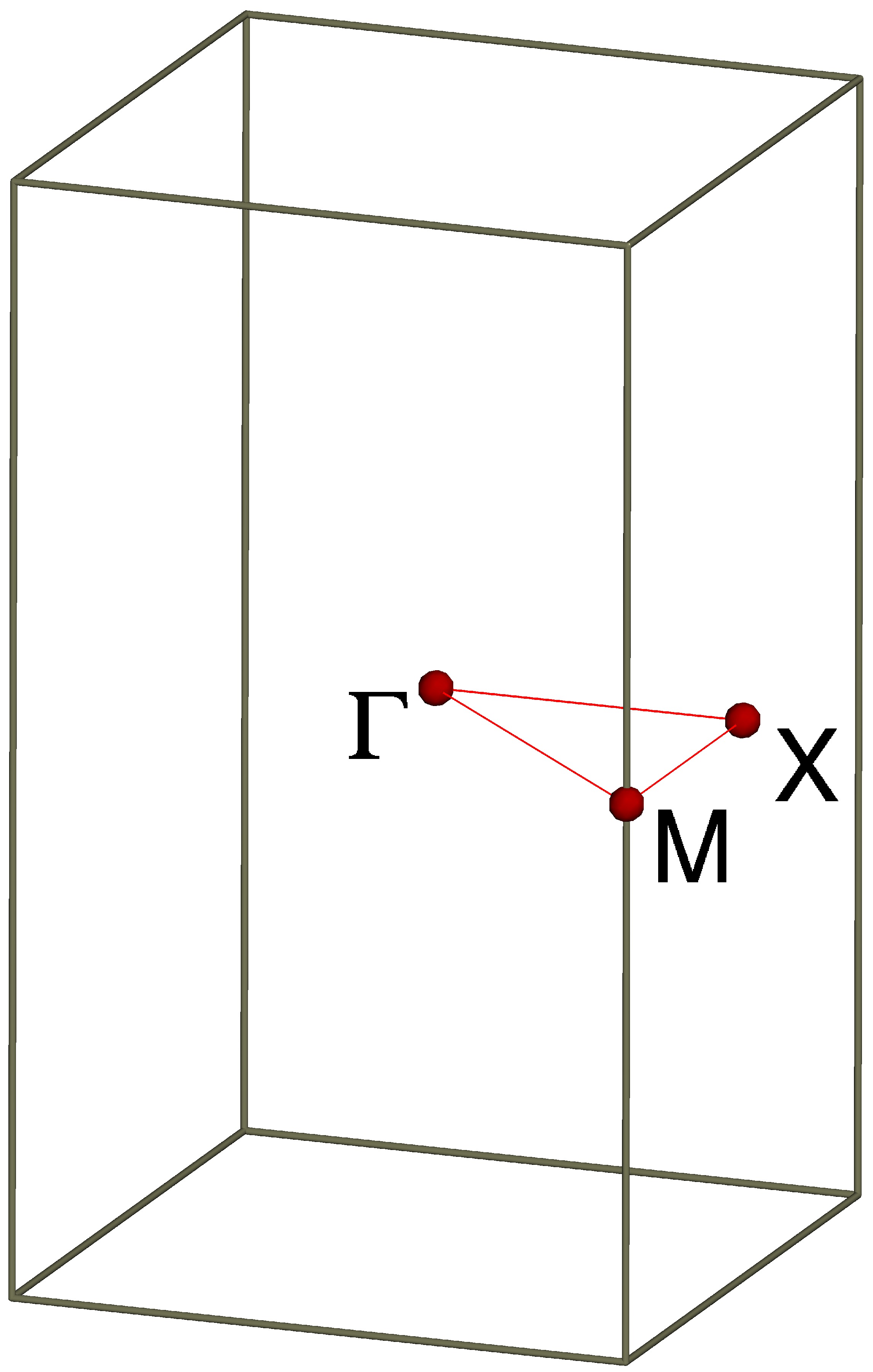}}}
\label{bands}
}
%\newline
\subfloat{
\put(10,165){b)}
\includegraphics[width=1\columnwidth]{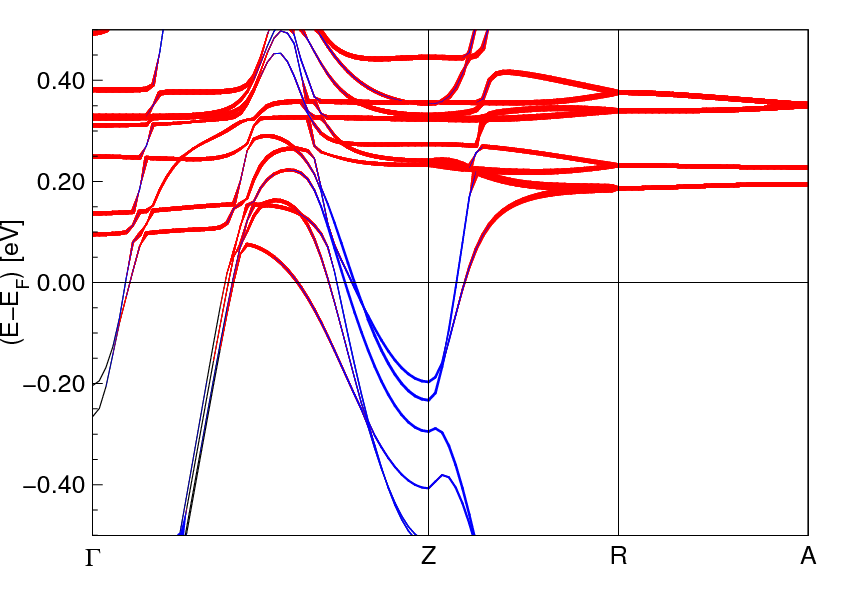}\llap{\raisebox{.5cm}{\includegraphics[height=2.5cm]{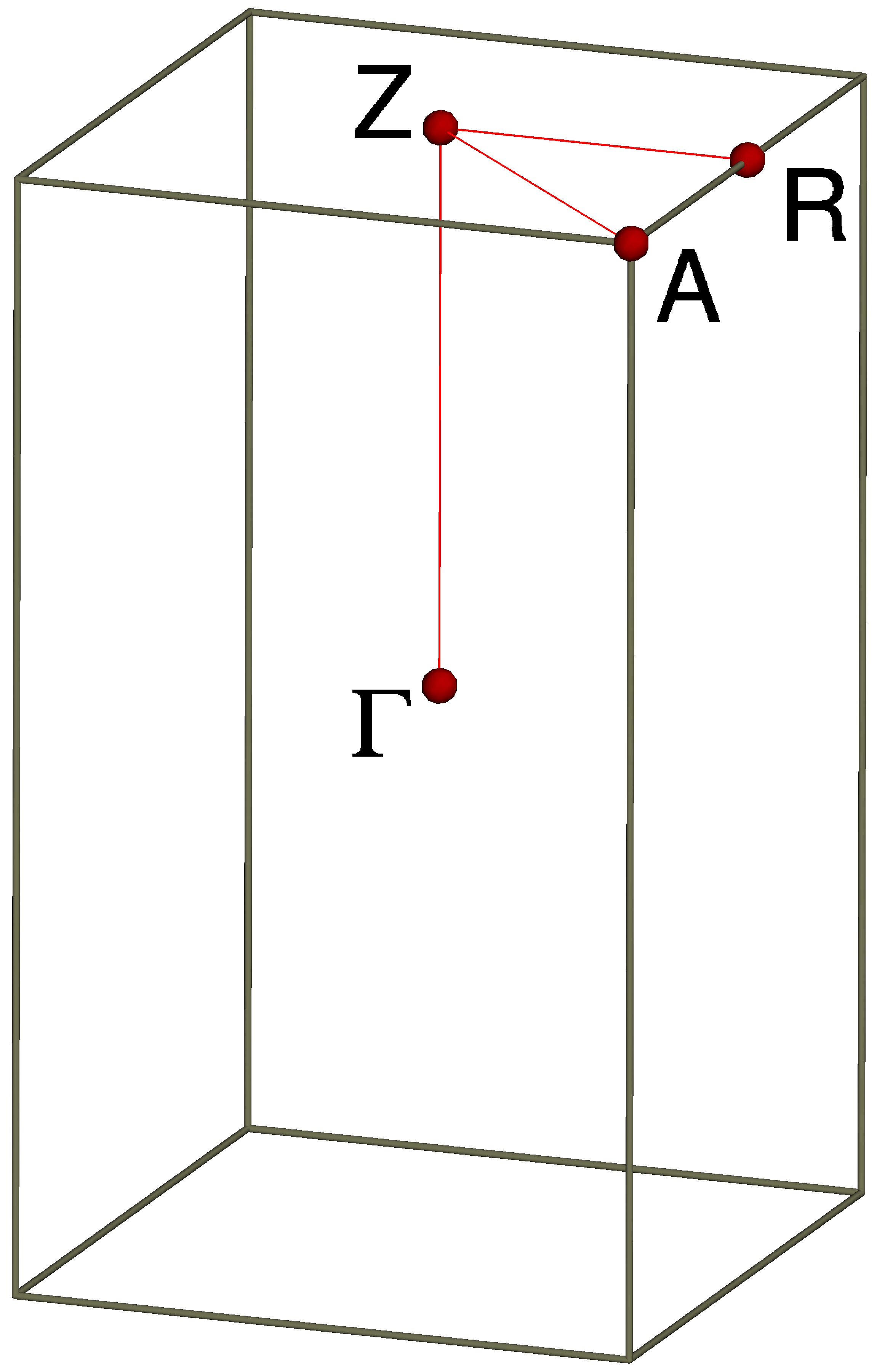}}}
\label{bands2}
}
%\subfloat{
%\put(10,165){c)}
%\includegraphics[width=1\columnwidth]{dft_dos2}
%\label{dos}
%}
\caption{a) Band structure for $\ce{TmB_4}$ from the DFT calculation, which shows dispersion in the $a$-$b$ plane. The inset shows the path taken in the BZ, $\Gamma$-$X$-$M$-$\Gamma$. Red indicates Tm $\ket{J=7/2}$ weights, while blue indicates dimer B $\ket{J=3/2, m_J=\pm1/2}$ weights. We have circled a region that shows strong $4f$-$2p_z$ hybridization, which occurs at the $M$ point, with a magenta circle. There are electron pockets around $\Gamma$, $X$, $M$, and along the $M-\Gamma$ path. b) Dispersion along the $c$-axis, where the inset shows the path taken in the BZ, $\Gamma$-$Z$-$R$-$A$. There are electron pockets around the $Z$ point, and hole pockets along the $\Gamma$-$Z$ path. $4f$-$2p_z$ hybridization can be seen along the $\Gamma$-$Z$ path and around the $Z$ point. There is some $\ce{5d}$-$\ce{2p_z}$ hybridization around the $Z$ point (not shown).}
\label{dft}
\end{figure*}

The band structure obtained using the DFT+U calculation is shown in Fig.~(\ref{dft}\subref{bands} and \subref{bands2}) where the vertical axis represents the energy of the Kohn-Sham eigenstate and the horizontal axis shows the position in the first Brillouin Zone. In Fig.~(\ref{dft}\subref{bands}), we focus on dispersion in the $a$-$b$ plane, where $\Gamma= (0,0,0)$, $X=(0, \frac{\pi}{b},0)$, $M=(\frac{\pi}{a}, \frac{\pi}{b}, 0)$, and a, b, and c are the lattice constants (where a = b in this case). The bands exhibit ``band-sticking" \cite{grouptheory} at high symmetry points due to the nature of the non-symmorphic space group, which contains a screw axis.

These plots show the rapidly dispersing bands associated with boron $\ce{2p}$ orbitals along with rather flat bands associated with $\ce{Tm}$ $\ce{4f}$ orbitals. We have circled two bands which show significant $4f$-$2p_z$ planar hybridization at the $M$ point.

The $M$ point is of particular interest due to significant hybridization between the $\ce{4f}$ and $\ce{2p_z}$ orbital that occurs there. At the $M$ point two boron dimer bands dip just below the Fermi level and there is a pocket of occupied states which are of $4f$-$2p_z$ hybrid character.  The $M$-point represents propagation along the in-plane diagonal in real space. It includes the four propagation directions which are parallel to the bonds between dimer boron atom pairs.

In Fig.~(\ref{dft}\subref{bands2}) we show $c$-axis dispersion, where the path taken in the BZ is $\Gamma$-$Z$-$R$-$A$, where $Z=(0,0,\frac{\pi}{c})$, $R=(0, \frac{\pi}{b},\frac{\pi}{c})$, and $A=(\frac{\pi}{a}, \frac{\pi}{b},\frac{\pi}{c})$. There is $4f$-$2p_z$ hybridization along the $\Gamma$-$Z$ path and around the $Z$ point. In addition, there is also some $\ce{5d-2p_z}$ hybridization around the $Z$ point (not shown). We have four hole pockets along $\Gamma$-$Z$, and two electron pockets around the $Z$ point.

\begin{figure*}
\subfloat{
\hspace{1cm}
\begin{overpic}[width=1.5\columnwidth]{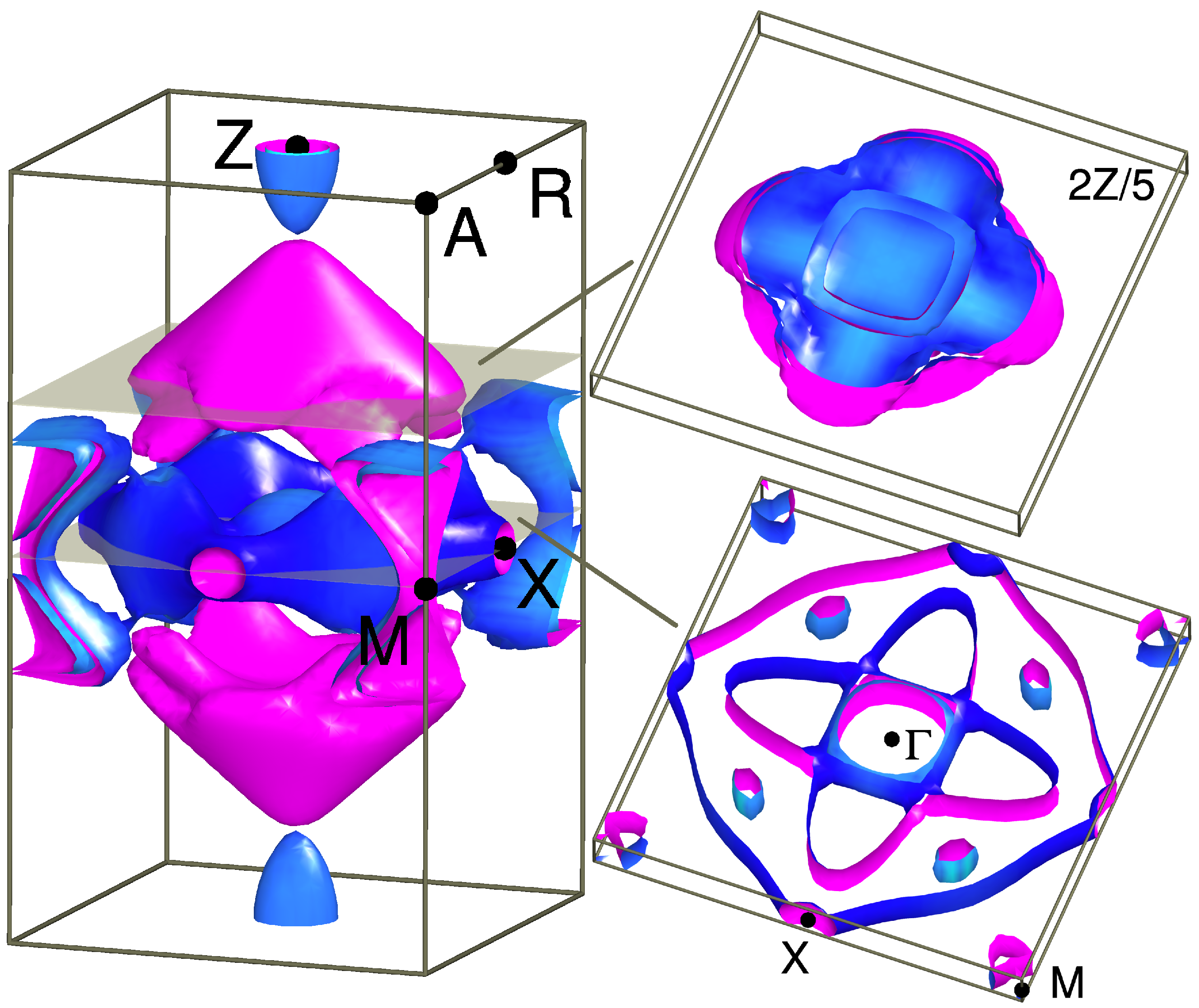}
 \put (-5,50) {\large$\displaystyle k_z$}
 \put (22, -0.5) {\large$\displaystyle k_y$}
 \put (45, 3.5) {\large$\displaystyle k_x$}
  \put (0,80) {a)}
 \label{full}
\end{overpic}
}
\newline
\subfloat{
\hspace{-1cm}
\begin{overpic}[width=.6\columnwidth]{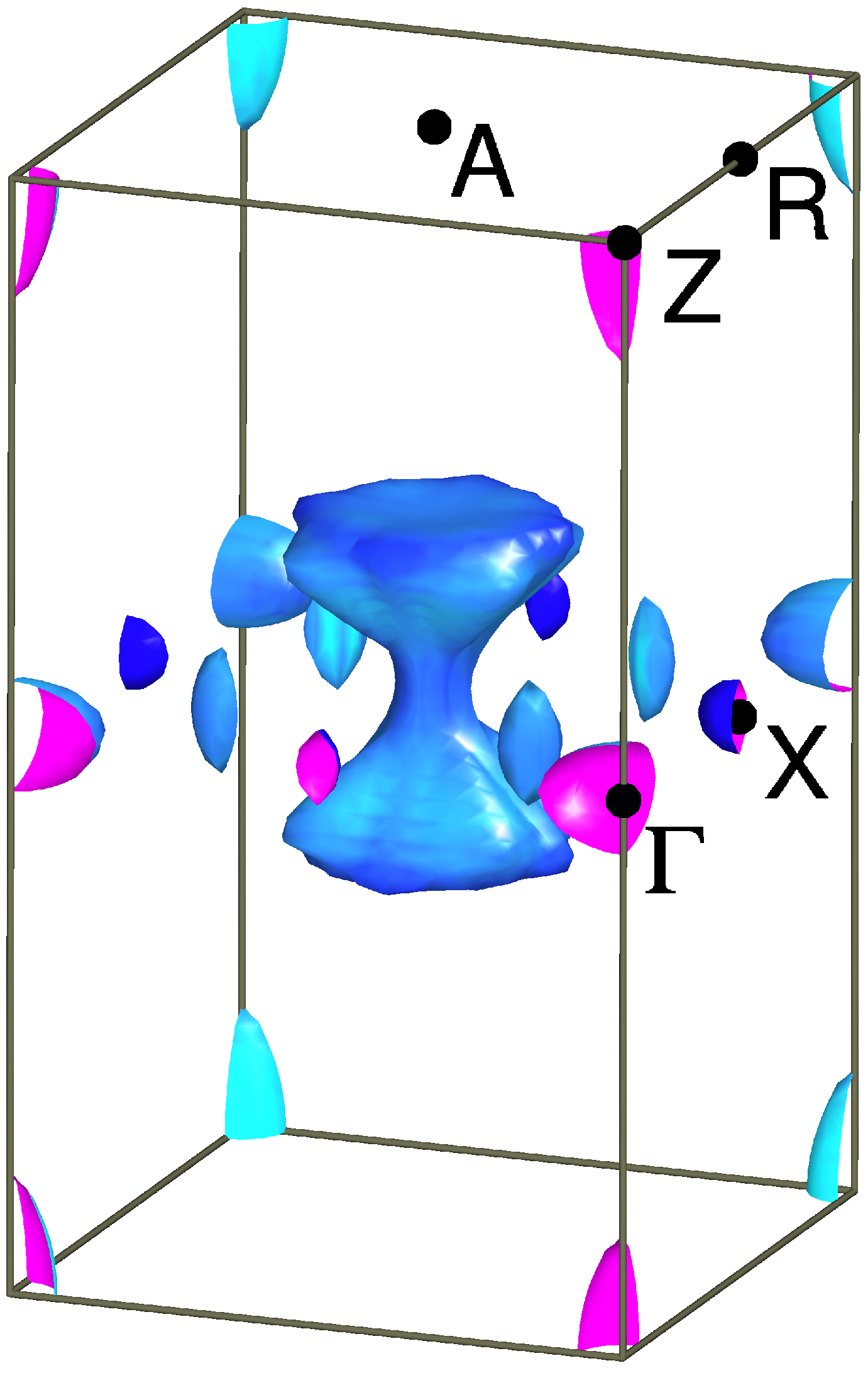}
 \put (-7,50) {\large$\displaystyle k_z$}
 \put (22.5,-2.5) {\large$\displaystyle k_y$}
 \put (55, 2) {\large$\displaystyle k_x$}
  \put (0,95) {b)}
 \label{m}
\end{overpic}
}
\subfloat{
\hspace{1cm}
\begin{overpic}[width=.6\columnwidth]{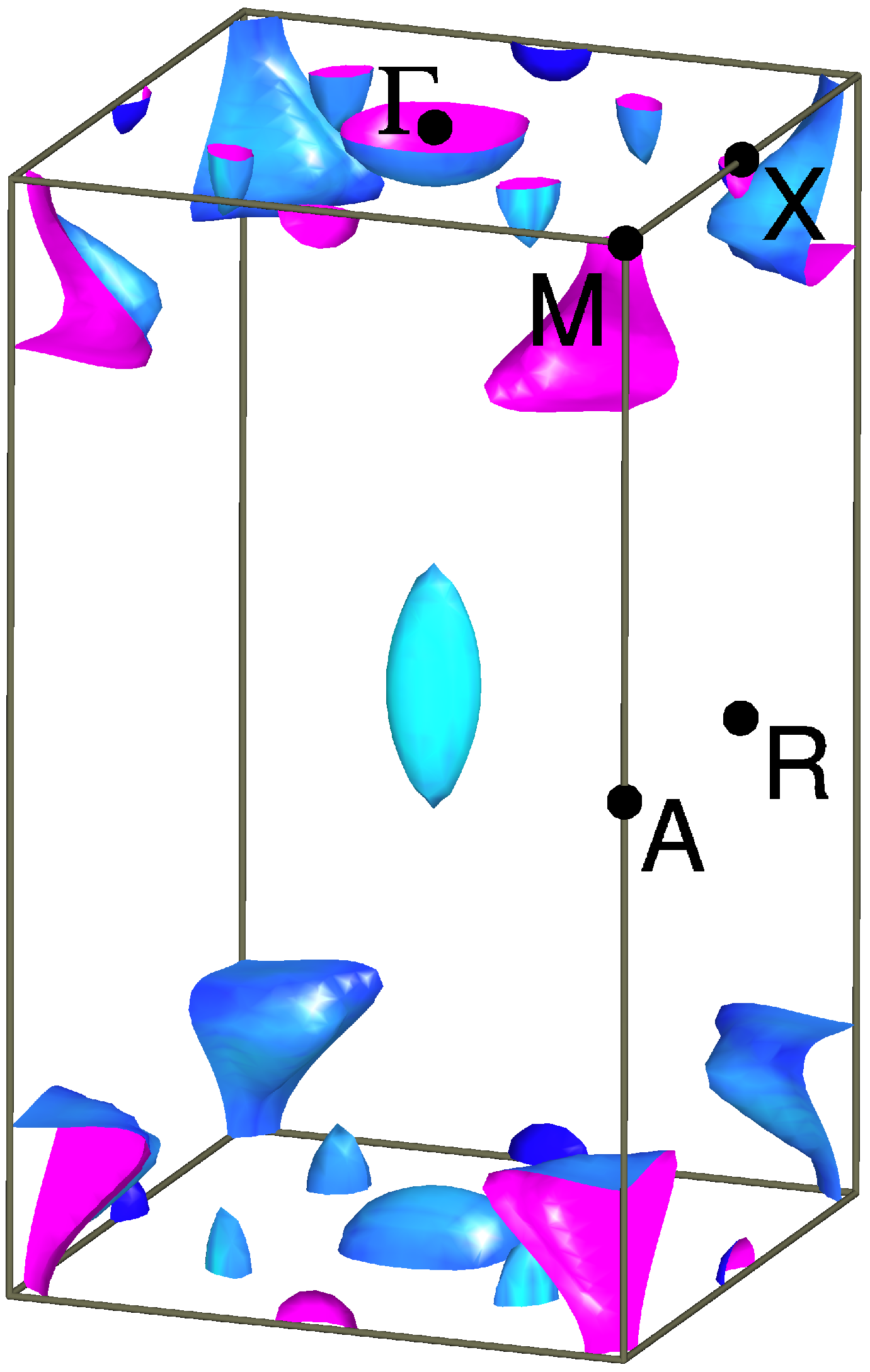}
 \put (-7,50) {\large$\displaystyle k_z$}
 \put (22.5,-2.5) {\large$\displaystyle k_y$}
 \put (55, 2) {\large$\displaystyle k_x$}
 \put (0,95) {c)}
 \label{z}
\end{overpic}
\begin{overpic}[width=0.5cm, height=3.1in]{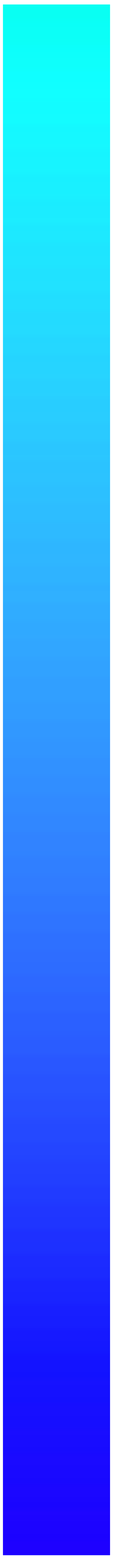}
  \put (8,97) {\large$\displaystyle 0.9 \cdot 10^{6} $ m/s}
  \put (8,0) {\large$\displaystyle 0.1 \cdot 10^{6}$ m/s}
\end{overpic}
}

\caption{Fermi surface from the DFT calculation, where the color indicates the magnitude of the Fermi velocity, as indicated by the color map.  In a), we show the full Fermi surface, centered at the $\Gamma$ point. Surfaces with an exterior blue gradient represent electron-like surfaces, while surfaces with magenta exteriors are hole-like surfaces: thus, the large (magenta) surface along the $\Gamma$-$Z$ path is a quasi 1-dimensional hole surface since it is flatter in the $x$-$y$ directions compared to $z$, while the others are electron surfaces. We have taken $x$-$y$ cross-sections of the electron-like surface at the $\Gamma$ point and of the hole-like surface at $2Z/5$, where the cross-sections have been given some height in the $z$ direction to show the curvature. In b) and c), we focus on a single hybridized $\ce{4f}$-$\ce{2p_z}$ band, centered at b) the $M$ point and the c) $Z$ point to better showcase the symmetry of the hybridized surfaces. The electron surface centered in c) shows quasi-2D conduction, with a planar effective mass ratio of $|\frac{m_{xx}}{m_{zz}}|$ = 0.067 near the tip of the surface. For the quasi-1D surface, we estimate an $m_{zz}$ of $0.46 \, m_e$ near $2Z/5$, while for the quasi-2D surface, we estimate an $m_{xx}$ of  $0.29 \, m_e$ near the tip. We also estimate carrier concentrations of 0.05, 0.04, 0.01, and 0.006 per cell for the four hole surfaces, and 0.006 and 0.004 per cell for the $Z$ point surfaces. The volume of the unit cell is $1.98 \cdot 10^{-28} \, \ce{m^3}$.}
\label{fermi}
\end{figure*}

Fig.~(\ref{fermi}\subref{full}), we show the full Fermi surface, where the large surface along $\Gamma$-$Z$ is a hole surface, and the rest are electron surfaces. The Fermi surfaces shown in previous measurements of the related rare-earth tetraboride $\ce{YB_4}$ \cite{yb4} can be clearly identified along the $\Gamma$-$Z$ direction. We can see examples of the aforementioned surface nesting around $M$ and $Z$. We have also taken $x-y$ cross-sections at the $\Gamma$ point and at $2Z/5$, to show how the surfaces nest. We can see that the cross-section at $2Z/5$ shows four hole surfaces, which can be grouped into two pairs, a larger pair and smaller pair. They exhibit quasi-1D behavior, as indicated by the relatively flat portions of the surfaces. For these surfaces, from largest to smallest, we have calculated the ratio of the effective masses at the $2Z/5$ point to be $\|m_{zz}/m_{xx}\| = 0.081, 0.076, 0.15,$ and $0.093$. We have also calculated the dHvA frequencies for the extremal cross-sectional areas in the plane defined by the $k_z$-axis for four major surfaces in Table \ref{dHvA}: the largest of the quasi-1D surfaces along the $\Gamma$-$Z$ path (the large magenta surface in \ref{fermi}\subref{full}, and the largest in the $2Z/5$ cross-section), the larger $M$ point surface (the surface in the midpoint of the edge of the side faces in \ref{fermi}\subref{full}, whose pair can be seen in \ref{fermi}\subref{m}), the larger $Z$ point surface (the surface at the top and bottom faces found in \ref{fermi}\subref{full}, whose pair can be seen in \ref{fermi}\subref{z}), and the large surface around the $\Gamma$ point, of which we can see the cross-sectional area in \ref{fermi}\subref{full}: 

\begin{table}[h!]
\centering

\begin{tabular}{c|c|c|c|c}

   & $\Gamma$-$Z$ & $M$ & $Z$ & $\Gamma$ \\
   \hline
Max $F$ (kT) & 0.75 & 0.44 & 0.058 & 1.40 \\
\hline
Min $F$ (kT) & - & 0.051 & - & - \\
\hline
\end{tabular}
\caption{dHvA frequencies for four major surfaces. The frequencies are in $10^3$ Tesla. The extremal cross-sectional areas are shown for planes defined by the $k_z$ axis. The four surfaces can be seen in Figure 4(a), where the $\Gamma$-$Z$ surface is the large magenta surface, the larger $M$ point surface is found on the midpoint of the edge of the side faces, the $Z$ point surface is found on the top and bottom faces, and the $\Gamma$ point surface is shown in the bottom cross-section, which touches the BZ boundaries. Only the $M$ point surface has a non-vanishing cross-sectional minimum.}
\label{dHvA}
\end{table}

In Fig.~(\ref{fermi}\subref{m} and \ref{fermi}\subref{z}), we show the Fermi surface for a single hybridized $4f$-$2p_z$ band centered at the b) $M$ point and the c) $Z$ point to show the anisotropy of the hybridization. There is a ``dumbbell" shaped surface centered at the $M$ point, and a ``football" shaped surface centered at the $Z$ point.

\section{Tight-Binding Models}

\begin{figure}[h!]
\includegraphics[width=1\columnwidth]{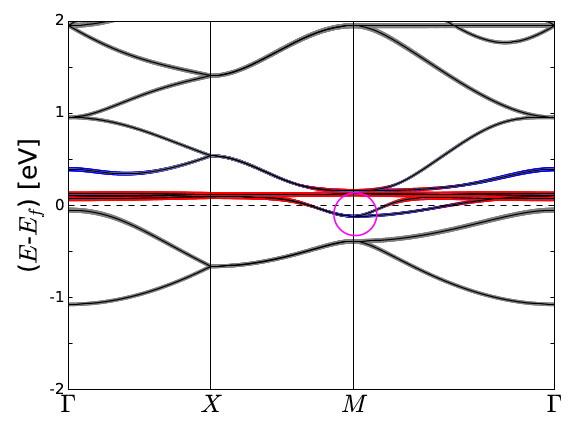}\llap{\raisebox{.75cm}{\includegraphics[height=2cm]{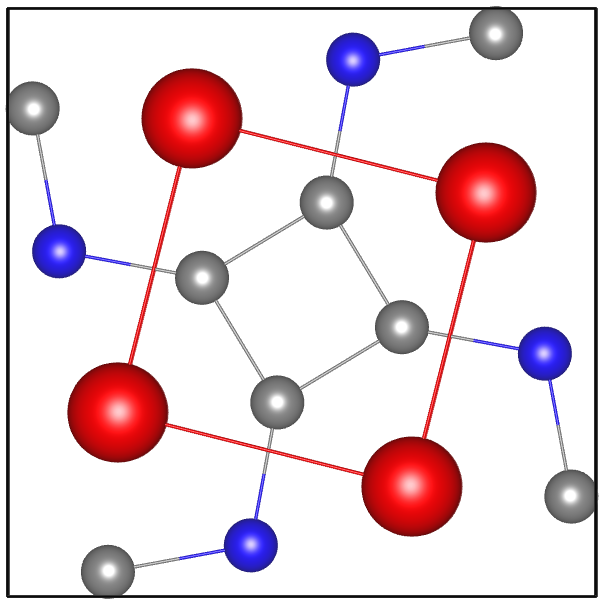}}}
\caption{Band structure of a simplified tight-binding model, where the apical boron has been removed. Tight binding band dispersion is calculated for $t=-1.0$ eV, $V=-0.04$ eV and $T=-0.01$ eV with site energies of $1.95$ eV for the planar B and $0.1$ eV for the Tm $\ce{f}$ orbital. Red lines reflect contributions from the $\ce{Tm}$ $\ce{f}$ eigenvector, while blue and gray indicate dimer and octahedral boron, respectively. The analogue of the two hybrid bands from the DFT calculation can be seen below the Fermi level ($E_F=0$) at the $M$ point, circled in magenta. The inset shows the unit cell for this simplified model.}
\label{tb_bands1}
\end{figure}

A goal of our tight binding calculations is to reproduce essential features of the bands using simplified models. We will focus purely on planar dispersion. Our first simplification is to ignore the apical boron atoms and to perform a tight binding calculation involving coupled planes of B and Tm atoms. This calculation will show very little $c$-axis dispersion, however, it will capture essential features of the $a$−$b$ plane dispersion and the coupling between Tm f-orbitals and boron $\ce{2p}$ bands. We include a single Tm $\ce{4f}$ orbital which couples to boron $\ce{2p_z}$ orbitals via the parameter $V$. In this calculation we use a single $\ce{2p_z}$ orbital for the boron atoms,  as the other $n = 2$ orbitals of the planar boron are involved in bonding \cite{lipscomb, yin, rareearth}, which are localized and non-interactive. This is what one expects based on the observed structure and 3-fold planar bonding, and it is confirmed by our DFT calculation, where the dimer boron $\ce{2p_z}$ orbital hybridizes with the $\ce{4f}$ level. The unit cell of this simplified model is shown in the inset to Figure~(\ref{tb_bands1}), where there are 4 $\ce{Tm}$ sites, eight equatorial or side sites (gray), and four dimer sites (blue).

We use the parameter $t$ to quantify the B to B hopping and $V$ for the coupling between the B $\ce{2p_z}$ orbital and the Tm $\ce{f}$ level. A small hopping parameter for in-plane $\ce{Tm}$ to $\ce{Tm}$ hopping, $T$, is also employed to give the Tm bands a little width. In Eq. \ref{tbham_1} and Eq. \ref{tbham_2}, $\ket{\phi}$ represents the planar boron $\ce{2p}$ wavefunction, while $\ket{\psi}$ represents the $\ce{Tm}$ $\ce{4f}$ wavefunction. In the first term, we have hopping between the boron and its 3 nearest neighbors. Next, we have coupling between the $\ce{Tm}$ wavefunction and its seven boron nearest neighbors, and finally, we have hopping between the $\ce{Tm}$ and its five near neighbors. 

\begin{multline}
\hat{H}_{TB} = t\sum_{<i,j>}\ket{\phi_i} \bra{\phi_j} + V\sum_{<i,j>}\ket{\psi_i}\bra{\phi_j} \\ + T\sum_{<i,j>}\ket{\psi_i}\bra{\psi_j} + hc
\label{tbham_1}
\end{multline}

In Fig.~(\ref{tb_bands1}) we show results for the band structure for this tight-binding model. $\Gamma$ is the center of the 1st BZ, i.e., the point at which $\vec{k}=\vec{0}$; $X$ is the side of the face and $M$ is the corner.  All of these represent in-plane dispersion. The $M$-point is of particular interest as considerable coupling between $\ce{4f}$ and $\ce{2p_z}$ states occurs there. This tight-binding calculation reproduces the pocket of occupied states of mixed $\ce{4f}$ and $\ce{2p}$ character in this region, where the analogue of the two hybrid bands highlighted in Fig.~(\ref{dft}\subref{bands}) can be seen just below the Fermi level ($E_F=0$), where the bands are degenerate from $X$-$M$ and split from $M$-$\Gamma$.

\begin{figure}[h!]
\includegraphics[width=1\columnwidth]{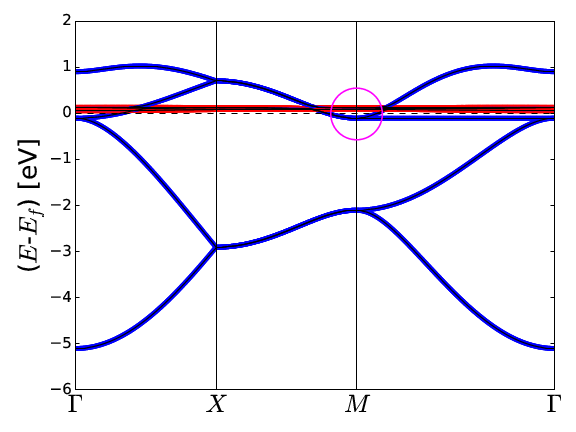}\llap{\raisebox{.75cm}{\includegraphics[height=2cm]{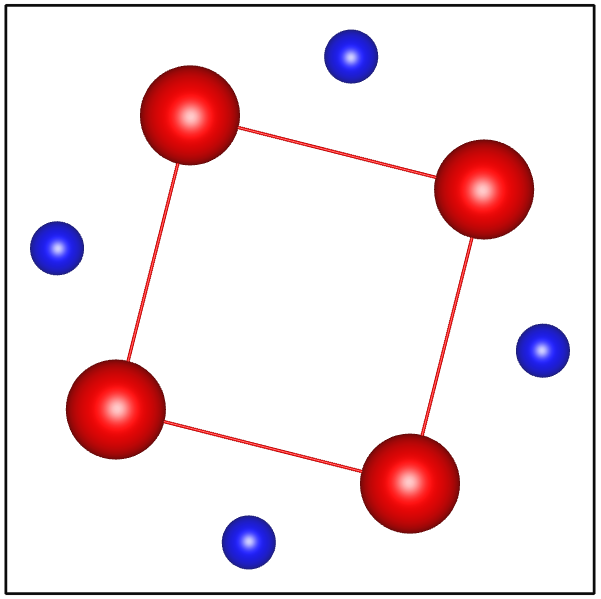}}}
\caption{Band structure of the further simplified tight-binding model, where the octahedral boron have been removed.  Tight binding band dispersion is calculated for $t=-1.0$ eV, $V=-0.04$ eV, $t_2=-0.75$ eV, and $T=-0.01$ eV with site energies of $-1.1$ eV for the planar B and $0.1$ eV for the Tm $\ce{f}$ orbital. The analogue of the two hybrid bands from the DFT calculation can be seen mixing at the $M$-point near the Fermi level ($E_F=0$), circled in magenta. The inset shows the unit cell for this further simplified model.}
\label{tb_bands2}
\end{figure}

Tight binding results from an even more simplified structure are shown in Fig.~(\ref{tb_bands2}). Because the $\ce{Tm}$ couples most strongly to the dimer boron, as inferred from our DFT results, we have also created a tight-binding model with just the Tm and dimer boron atoms, as illustrated in Fig.~(\ref{structure}\subref{dimer_lattice}) and in the inset to Fig.~(\ref{tb_bands2}), where the unit cell contains four $\ce{Tm}$ sites and four dimer sites. This calculation shows the two bands of mixed $\ce{f}$-$\ce{p}$ character at the $M$ point near the Fermi level ($E_F=0$) as in the previous TB calculation and in the DFT calculation. We have incorporated an additional parameter, $t_2$, which connects the pairs of borons. In Eq. \ref{tbham_2}, we have nearest neighbor hopping between the pairs of dimers, coupling between the $\ce{Tm}$ and its three dimer nearest neighbors, hopping between $\ce{Tm}$ sites, and a next nearest neighbor hopping which connects the isolated dimers, which is necessary for dispersion in this model. In these models, one could duplicate the pocket at the $X$ point found in the DFT band structure as in Fig.~(\ref{bands}) with a complex hopping parameter on the $\ce{Tm}$ lattice, but we have omitted this for simplicity.

\begin{multline}
\hat{H}_{TB} = t\sum_{<i,j>}\ket{\phi_i} \bra{\phi_j} + V\sum_{<i,j>}\ket{\psi_i}\bra{\phi_j} \\ + T\sum_{<i,j>}\ket{\psi_i}\bra{\psi_j} + t_2\sum_{\ll i,j \gg} \ket{\phi_i} \bra{\phi_j} + hc
\label{tbham_2}
\end{multline}

\section{Simplified Crystal Field Physics of the $\ce{Tm}$ ion in $\ce{TmB_4}$ \label{CF}}
In order to build up our understanding of the Kondo type model,   and further a suitable  interacting  model of the spins, we need to make several approximations. We will simplify matters from the formally exact but technically  formidable periodic Anderson lattice model, and view the $\ce{Tm}$ initially as an impurity embedded in the $\ce{B}$ lattice. In a unit cell one has four inequivalent $\ce{Tm}$ ions, but each has the same local environment that is rotated relative to the others.  We consider any one $\ce{Tm}$ as our impurity and describe its CF level structure next.
 The $\ce{Tm^{3+}}$ ionic state has a $\ce{f^{12}}$ configuration, leading to an even number of electrons.  It  thereby evades the Kramers degeneracy that  is responsible for Ising like behavior in many clean examples,  such  as  the  case of $\ce{Dy^{3+}}$ ions in $\ce{DyBa_2Cu_3O_{7-\delta}}$ \cite{dirken,ramirez}, which is an excellent realization of the famous 2-d Ising model  of  Onsager\cite{Onsager}. 
Therefore  the origin of the  Ising like behavior reported in experiments, with $m \sim \pm 6$ requires some explanation. Towards this end we present a simple crystal field  (CF) theory calculation using a point charge crystal field  model \cite{Hutchings}.  The results of this model need to be taken with caution  in a metallic system, since the  charges are smeared in a metallic system, as opposed to an insulator. 

\begin{figure*}
\begin{center}
\includegraphics[width=\columnwidth]{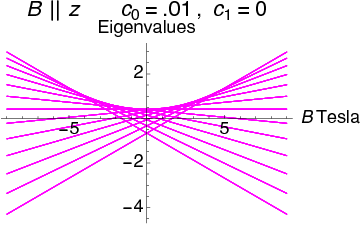}
\includegraphics[width=\columnwidth]{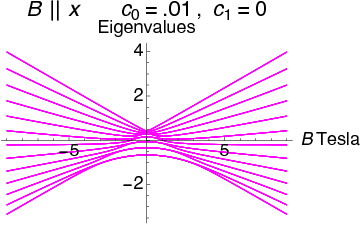}
\includegraphics[width=\columnwidth]{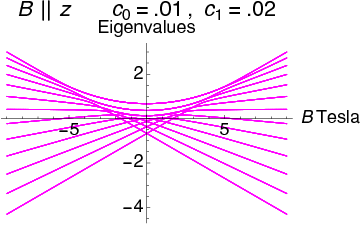}
\includegraphics[width=\columnwidth]{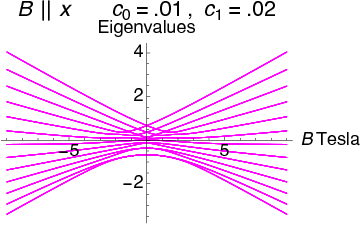}
\end{center}
\caption{The full spectrum of $H_{CFZ}$ the crystal field plus Zeeman Hamiltonian with magnetic field along $z$ axis (panels at left), or along $x$ axis (panels at right)  from \disp{vcf-Zeeman} with  $c_0=.01$ meV,  and $c_1=0$ meV (top-left and top-right) or $c_1=0.02$ meV (bottom-left and bottom-right). The top-left panel illustrates the effective Ising nature of the Tm moment in the $z$ direction  with a linear evolution of the lowest energy with field $B || z$ over a wide range. We see here the necessity for  choosing  $c_0>0$, inverting its sign also inverts the spectrum and the ground state then would be non Ising-like for fields $B_Z \lessim 5$ Tesla.
The top-right panel shows that a field along the $x$ axis $B||x$, the lowest level  is not linear, but for fields around $6$ T the linearity is recovered with a  slope (i.e the magnetic moment) $\sim 6 \mu_B$. The two lower panels have a non zero value of the transverse terms, i.e.  $c_1 \sim 0.02$ meV. This does not create a noticeable gap in the level crossing at zero field in the z direction- while a larger value would do so. It thus appears that we can ignore $c_1$ for many (but not all) purposes, as explained in the text.}
\label{CF-Zeeman}
\end{figure*}

 We make one important change from standard CF theory, the sign of the crystal field energy is taken to be  opposite of the usual  one valid in an insulator. An f-electron on $\ce{Tm^{3+}}$ will be taken to experience an attractive (rather than repulsive) force from  the B sites. The repulsive interaction sign convention inverts the spectrum and gives  a vanishing moment, clearly a wrong result. This can be seen from the spectrum of the ionic Hamiltonian in \figdisp{CF-Zeeman}, where the opposite sign simply inverts the spectrum.

Flipping the sign of the CF interaction in metallic systems has a well established precedent. In the case of rare earths (including $\ce{Tm}$) in noble metal ($\ce{Au}$ and $\ce{Ag}$) hosts, the significant work of Williams and Hirst in \refdisp{Hirst-2}  shows that the observed moment requires such a flipped sign.  Flipping the sign in  \refdisp{Hirst-2} (see also \refdisp{mulak-book} pages 171-173)  was ascribed to the presence of 5-d electrons in the rare earth, which are argued to form a  virtual bound state at the Fermi level. The polarizable nature of this virtual bound state is taken to allow for  such a flipping. The DFT calculation shows an occupied 5d state, along with many negative hopping integrals for the corresponding $\ce{Tm}$ and $\ce{B}$ Wannier functions, which would support this argument. 

An independent  argument made below invokes the weakly electronegative nature of the boron atom. It is not completely clear as to  which of the two rather qualitative  arguments is ultimately responsible for the final effect, but since the latter is  simple enough, we present it anyway.
We will  choose  the point charges at the boron atoms to be  {\em positive}, corresponding to an ionized state lacking some electrons. This change is motivated by the knowledge that in the metallic state being  modeled, the boron atom donates its electrons towards band formation. It is thus  very far from the ionic limit in an insulator, where the atom  might be  imagined  to have captured some electrons. This  contrasting situation is realized by the fluorine atom in $\ce{LiHoF_4}$, or the oxygen atom in stoichiometric high $T_c$ parent compound $\ce{La_2CuO_4}$. The relevant atoms $\ce{F}$ and $\ce{O}$  have a  high  electron affinity, and therefore grab one and two electrons respectively in the solid, and   may be thus be visualized as negatively charged. Thus we might expect  that the boron atom, with its considerably smaller electron affinity \refdisp{electronaffinity-boron}, plays a role closer to that of an anion, rather than a cation, in order to reconcile the data on $\ce{TmB_4}$.

For this purpose we consider the total electrostatic potential at the location of a $\ce{Tm}$ ion, produced by its neighborhood of 18 boron atoms located as follows:  the four apical sites, \cite{apical}, the six dimer sites \cite{dimer}, where two of the borons are a larger distance away from the $\ce{Tm}$ site, and the eight ``side" boron which form the equator of octahedra \cite{side}. We can visualize  these sites as in Fig.~(\ref{local}).
%\end{widetext}
Assuming  an attractive Coulomb interaction between the electron and each boron atom, the Coulomb energy can be expanded at the $\ce{Tm}$ site, assumed to be the origin, and leads to 
\beq
V_{CF}= V_0 + V_1 (x+y) +V_2 (x^2+y^2-   2 z^2 -  \alpha \; x y) + O(r_\alpha^3),  \label{vcf}
\eeq
where $V_0,V_1$ and $V_2= 0.0432942$ are   constants, $\alpha =5.39965 $  and the neglected terms are of third and higher order in the components of the vector $\vec{r}= \{x,y,z\}$ locating the $\ce{Tm}$ ion. If we choose a repulsive interaction, as appropriate for the case of say $\ce{LiHoF_4}$ or stoichiometric $\ce{La_2CuO_4}$,  the Coulomb interaction would be chosen as repulsive. Thus flipping the sign of our final result would give the repulsive case.

Observe that this series contains odd  terms  in the coordinates, this is due to  the lack of inversion symmetry about the $\ce{Tm}$ ion. One implication is that we expect to see optical transitions between different crystal field levels. It would be interesting to pursue this using optical methods, especially since $\ce{Tm^{3+}}$ ions are studied in  infrared laser materials \cite{infrared-lasers, infrared-lasers2}. 
 From \disp{vcf} we can construct an effective local spin Hamiltonian  by using the Wigner Eckhart theorem (W-E). This procedure can be automated for ions having a simple symmetry, as in the Stevens effective Hamiltonian theory \cite{Hutchings}.  In the present case the symmetry of $\ce{Tm}$ site is very low and hence it is more useful to build up our understanding  from the basics. The manifold of angular momentum $|J|=6$ states, arising from the Hunds rule as $\vec{J}= \vec{L}  + \vec{S}$, with $|L|=5$ and $|S|=1$,  is the basis for the representation of all vector operators.
  The W-E theorem helps us to replace the components  of the vector $x, y,z$ by operators proportional to  $J_x,J_y,J_z$.  More generally  we set
\beq
r^n_\alpha \to M_n J^n_\alpha
\eeq
where $M_n$ is the W-E reduced matrix element. Due to parity of the $\ce{4f}$ matrix element, only  terms in $r^n_\alpha$ with even $n$ survive \cite{parity}. Therefore  to second order, we  write the effective local  Hamiltonian as
\begin{eqnarray}
H_{CF}/c_0 & =&   ( J_x^2+J_y^2 - 2 J_z^2 ) - \frac{\alpha}{2} (J_x\,J_y+J_y\,J_x) \label{vcf2} \\
&=& J(J+1) - 3 J_z^2  + i   \frac{\alpha}{4} \left(J_+^2 - J_-^2   \right). \label{vcf3}
\end{eqnarray}
where we used a symmetrization rule for non commuting operators to express the $xy$ term  in \disp{vcf} in terms of components of $J$ in  \disp{vcf2}. 
Here $c_0$ ($=V_2 M_2^2$) is a constant that lumps together the reduced matrix element, shielding factors  and other details - it is usually best to determine it from experiments. This theory is missing the fourth and higher order terms in $r_\alpha$, these need not be small but one expects  that the low order theory given by \disp{vcf3}  provides the correct starting point for discussing  the CF splitting of the $J=6$ level.

In \disp{vcf3} we see that if $\alpha=0$, the leading term in the Hamiltonian has an exact Ising like symmetry leading to a degenerate   minimum at $J_z= \pm 6$. However since $\alpha$ is estimated above to be nonzero,  this symmetry cannot be exact, and we must examine the solution further.  If we use 
the raising and lowering terms with coefficient $\alpha$ as a perturbation of the leading term, the degeneracy is lifted to the sixth order, and the resulting states will be two distinct linear combinations of the  maximum  $|m|=  6$ states,  with amplitudes for lower $|m|$. We can also calculate the eigenstates numerically, a simple calculation yields the two low lying solutions 
\begin{widetext}
\barray
E_a/c_0& =& -78.5199,\;  \nn \\
\Psi_a &=& \{-0.559111, 0, - 0.319154 \, i, 0, 0.245299, 0, 
  0.225227 \, i, 0, -0.245299, 0, 
  - 0.319154 \, i, 0, 0.559111\}  \nn \\
E_b/c_0& = & -75.4681, \nn \\
 \Psi_b& =& \{-0.640364, 0, 
  - 0.276433 \, i, 0, 0.116273, 0, 0, 0, 0.116273, 0, 
  0.276433 \, i, 0, -0.640364\},
\earray
\end{widetext}
where the wave functions are expressed in the basis $|m \rangle, \; m=6,5,\ldots -5,-6$. As expected from the perturbative argument, the two states  are   distinct  linear combination of 
$ |m\rangle$ states with $m=6,4,2,0,-2,-4,-6$, the skipping of odd $m$ is  as required by the form of the perturbation term $ i   \frac{\alpha}{4}  \left(J_+^2 - J_-^2   \right)$. The signs of the coefficients imply that the two states are derived from the sum and difference of the two degenerate states $m=\pm 6$. Each state also has a vanishing expectation of $J_z$ ,  and is orthogonal to each other. Thus, in the absence of a magnetic field, the two lowest states,
split off from higher energy states by a gap,
 are 
not exactly degenerate, but rather are linear combinations of the two $m= \pm 6$ states of the Ising model.

%\FloatBarrier
To understand the behavior in a magnetic field we add the Zeeman energy to \disp{vcf3} and consider the Hamiltonian
\begin{eqnarray}
H_{CFZ} &=& H_{CF} - g \mu_B h J_z \nn \\
 & =& c_0 \left(J(J+1) - 3 J_z^2 \right) + i \, c_1 \left(J_+^2 - J_-^2   \right) - g \mu_B \vec{B}.\vec{ J}  \nn \\ \label{vcf-Zeeman}
\end{eqnarray}
where $c_1= c_0 \frac{\alpha}{4}$ and $g=\frac{7}{6}$ from the Lande rule.  We plot the eigenvalues of $H$ versus $g \mu_B h$ at two values of $c_0,c_1$  in \figdisp{CF-Zeeman}. Constraints on the value of $c_0$ follow from the experimental observation in \refdisp{Siemensmeyer} that $TmB_4$ displays a moment of $\sim 6 \mu_B$ for a field along the z (i.e. c) axis. In the x direction the full moment $\sim 6 \mu_B$ is regained only for a field $B \gssim 6$ T.  At $c_0 \sim .01$ meV the condition on the magnetic moments  is satisfied, while smaller (larger) values of $c_0$ make the recovery in the transverse direction occur at lower (higher) values of the $B$ field.  The choice of  $c_1$ is less stringently constrained, from the estimates on $\alpha$ we expect $c_1\lessim.02$ meV.  The value of $c_0\sim.01$ meV is too  small to create a gap near $B\sim 0$  between the two levels that cross there, and hence we see that the moment of $Tm$ can be modeled well by an effective Ising model. The small but non-zero value of $c_1$ plays the role of mixing the 
 $m=6$ components  with lower $m=4,2,0 \ldots$ as discussed below, this feature is essential for the emergence of a Kondo-Ising model.

\section{Kondo Ising Model for $\ce{TmB_4}$ \label{Kondo}}

We next formulate a minimal model for describing $\ce{TmB_4}$ following the method indicated in \refdisp{aligia, aligia2} and \refdisp{Hewson} in the context of the mixed valent  compound  $\ce{TmSe_2}$. We first set up a local  type Anderson or Hirst type model  that incorporates the crystal field split Tm levels, and the boron bands that hybridize with these.
The $\ce{Tm}$ ion is considered to be in its ground (excited) state with 12 (13) electrons.  The relevant  $4f^{12}$ level has eigenstates given by $|J_0 M_0\rangle$ with $J_0=6$ and $-6\leq M_0 \leq 6$. The excited  state  $4f^{13}$ has eigenstates given by $|J_1 M_1\rangle$ with $J_1=\frac{7}{2}$ and $- \frac{7}{2}\leq M_1 \leq \frac{7}{2}$. We will use these ranges for the symbols $M_0$, $M_1$ and their primed versions below.
The ground state of the  ionic Hamiltonian of the  was discussed in Section (\ref{CF}). We use a similar scheme for the first excited state with angular momentum $J_1$ and   write a generalized ionic Hamiltonian:
\begin{widetext}
\begin{eqnarray}
H_{ion} &=& \sum_{M_0} E_{12}(M_0)  |J_0 M_0 \rangle \langle J_0 M_0|+ \sum_{M_1} E_{13}(M_1)  |J_1 M_1 \rangle \langle J_1 M_1| \nn \\
&&+  i c_1 \left\{\;  |J_0 M_0 +2\rangle \langle J_0 M_0| - hc \right\}+ i c'_1  \left\{\;  |J_1 M_1 +2\rangle \langle J_1 M_1| - hc \right\}, \label{Hion-1}
\end{eqnarray} 
\end{widetext}
and
\begin{eqnarray} 
E_{12} &=& c_0 (J_0(J_0+1) - 3 M_0^2)  \label{Hion-2}\\
E_{13} &=& c'_0 (J_1(J_1+1) - 3 M_1^2) \label{Hion-3}
\end{eqnarray} 
In this formulation, $c_1$  plays the role of quantum corrections to the otherwise diagonal model, which contains a preference for the largest magnitude $M_0=\pm6$. As discussed in  Section (\ref{CF}), the role of $c_1$ is to mix states $M_0=\pm4, \pm2 \ldots$ into these states, and to  slightly lift the degeneracy between $M_0 = \pm 6$. 
Clearly  terms with $c_1'$ play a similar role in the higher multiplet. We could ignore $c_1$ and proceed with the pure Ising model, but we will see below that these mixing terms play an important role in producing the Kondo-Ising model. 
Our full  Hamiltonian $H_{total}$ has two terms in addition to $H_{ion}$. First we have  the band energy
\beq
H_{band}= \sum_{k j m} \varepsilon_{k j m} C^\dagger_{ j m}(k) C_{jm }(k), \label{Hband}
\eeq  
where we have projected the conduction electron states into angular momentum resolved states about the $\ce{Tm}$ atom assumed to be at the center, i.e. $C^\dagger_{ \sigma}(\vec{k})=\sum_{j m} \langle \vec{k} \sigma| k j m\rangle C^\dagger_{j m}(k)$. Second we have a mixing term between the conduction and the $f$ electrons. To write this we  temporarily forget about the CF terms-  thus  assuming that angular momentum is conserved, we write:
\beq
H_{mix} = \sum_{j= \frac{5}{2}}^{ \frac{7}{2} } \sum_{m_j =-j}^j\sum_k V_k^{(j)} (f^\dagger_{j m} C_{j m}(k) + hc), \label{Hmix}
\eeq
where $V_k^{(j)}$ is a hybridization matrix element.
We have denoted the angular momentum resolved f-level Fermion as $f_{jm}$, here the allowed values  $j=\frac{5}{2}, \frac{7}{2}$,  are found by adding $L=3$ from the f level and the spin half of the electron. Since the $f^{13}$ and $f^{12}$  states are analogous to bound complexes of electrons, it is convenient to   rewrite  \disp{Hmix} as
%\begin{widetext}
\begin{eqnarray}
H_{mix}&=&  \sum_{k,j,m_j} \sum_{M_0,M_1}  V_k^{(j)} \langle J_0 M_0 j m_j|J_1 M_1\rangle \nn \\
&&\times \left( C^\dagger_{j m_j}(k) \; |J_0 M_0\rangle \langle J_1 M_1|+ hc\right),
\end{eqnarray}
%\end{widetext}
where the Clebsch-Gordon coefficient $\langle J_0 M_0 j m_j|J_1 M_1\rangle $ enforces
$M_0+m_j=M_1$.  The allowed ranges of the variables $M_0,M_1,j,m_j$ are summarized as
$|M_0|\leq 6$, $|M_1| \leq \frac{7}{2}$, $j=\frac{7}{2}$ or $j=\frac{5}{2}$, and $|m_j| \leq j$.  
 We should view $|J_0 M_0\rangle$  as a bound complex of the 12-f electrons, and similarly the state $|J_1 M_1\rangle$. The   transition between these by adding an f electron is expected to have a small matrix element, which is absorbed in to the symbol $V_k^{(j)}$. The total Hamiltonian is thus 
\beq
H_{total} = H_{ion} + H_{band} + H_{mix}. \label{Htot}
\eeq
We can find the analog of the Kondo model from \disp{Htot} by using the standard Coqblin-Schrieffer transformation \cite{coqblin-schrieffer}.  We may symbolically write it  as  $ H_{eff} = H_{mix}.\frac{1}{\Delta}. H_{mix}$ where  the  intermediate state energy $\Delta \sim E_{13}-E_{12}$. We  thus
 write down the effective low energy model-the Kondo-Ising model for $|J|=6$   as: 
 \begin{widetext}
 \begin{eqnarray}
 H_{KI}  &=& - \sum   J_K  \left( C^\dagger_{k' j' m'_j} C_{k j m_j} \right) |J_0 M_0'\rangle \langle J_0 M_0 | \; \times  \delta_{m_j+M_0,M_1} \delta_{m'_j+M'_0,M_1} + H_{CFZ} ,  \nn \\
 H_{CFZ}&=&-  3    c_0  \sum_{M_0} M_0^2   |J_0 M_0 \rangle \langle J_0 M_0| +   i c_1 \sum_{M_0} \left( \;  |J_0 M_0 +2\rangle \langle J_0 M_0| - h.c. \right) - g \mu_B \vec{B}. \vec{J},  \label{KI}
 \end{eqnarray}
 \end{widetext}
 where we added the  CFZ  term from \disp{vcf-Zeeman} after discarding  a constant. We expect $c_0 \sim .01$ meV and  $c_1 \sim .02$ meV.
Given the allowed range of the variables, we see that the  permissible  transitions are governed by $M_0'-M_0=m_j-m'_j$.  Its maximum magnitude  is 7 from the range on the right hand side $|m_j|, |m'_j|\leq \frac{7}{2}$. Hence  the states $M_0=6$ and $M_0=-6$ cannot be connected by \disp{KI} directly. Therefore the model cannot be immediately mapped into a simple effective Kondo-Ising model, where all transitions between allowed $M_0$'s are possible. However  we recall that the mixing term (i.e. $c_1$ etc.) in $H_{CFZ}$ allows a mixing between $|M_0|=6$ and lower values in steps of 2. This model can be further mapped into the Ising doublet manifold by using higher order degenerate perturbation theory in $H_{KI}$ and $c_1$. In summary it requires a careful consideration of the various mixing terms to recover the Kondo-Ising  model with a full range of angular momentum.   

  Within a perturbative approach in  $c_1$ (i.e. $\alpha$)   it is clear that an effective Kondo Ising model emerges with the full range of allowed transitions.  We  also see from a standard argument (see \refdisp{Hewson})  that the elimination of the conduction electrons in \disp{KI} leads to a long ranged   RKKY interaction of the Ising type. Such a model is of much interest theoretically \cite{Dublenych, Pinaki-plateaux, huang-theory}  and could also be invoked at the lowest order to understand the experiments on the magnetization plateaux in this system \cite{Ising-plateaux}. We should note that  quantum effects, coming in at higher orders in $\alpha$  (i.e. $c_1$) might be relevant in obtaining a good  
   understanding of  the anisotropic magnetic response.

\section{Conclusion}

In this paper we examine the electronic and magnetic characteristics of $\ce{TmB_4}$.  Our results from an ab-initio density functional theory approach provide insight into simplifications of the lattice in reduced tight-binding models. In the $a$-$b$ plane, $4f$-$2p_z$ hybridization around the $M$ point is a strong contributor to the in-plane conductivity. We have also found effective 1-D conduction along the $c$-axis, indicated by relatively flat segments of the Fermi surface which can be found along the $\Gamma$-$Z$ path. For the largest of these surfaces, we have calculated the effective mass ratio to be $\|m_{zz}/m_{xx}\| = 0.08$.

We then examined the local structure, building a simplified crystal field model containing the essential physics and which led to a description of the pseudo-Ising nature of the ground state, in which the degeneracy has been slightly lifted by off-diagonal terms originating from the lack of inversion symmetry at the $\ce{Tm}$ site.

From this point of departure, we constructed a Hamiltonian consisting of a band term, an ionic term, and a mixing term. Using the Coqblin-Schrieffer transformation, we constructed an effective low-energy model, an effective  Kondo-Ising type model, in which the (quantum) off-diagonal terms in the crystal field Hamiltonian are necessary  in recovering the full range of angular momentum. At low magnetic  fields $|B|\lessim 10$ T  the Ising approximation of the Kondo model is validated for our choice of the anisotropy constant $c_0$. Further experimental studies in transverse fields should help to refine the values of constants $c_0$ and $c_1$ in \disp{KI}.  It seems plausible that the elimination of the conduction electrons would lead to an RKKY-type Ising model with some quantum corrections, and thus refine the usual starting point of  studies on the magnetization plateaux \cite{Pinaki-plateaux}. Using the experimental distance and moment, we estimate that the long ranged dipolar exchange  J is $\sim$ 0.74 K at the nearest neighbor separation, and hence this needs to be added to the RKKY type interaction for obtaining the magnetic behavior at low T.
 
\section{Acknowledgements}

The work at UCSC was supported by the U.S. Department of Energy (DOE), Office of Science, Basic Energy Sciences (BES) under Award \# DE-FG02-06ER46319. This work also used the Extreme Science and Engineering Discovery Environment (XSEDE), which is supported by National Science Foundation grant number ACI-1053575. We thank Alex Hewson, Sreemanta Mitra,  Christos Panagopoulos, Art  Ramirez and Pinaki Sengupta  for helpful discussions on the project. We would also like to thank Jos\'{e} J. Baldov\'{i}, Alejandro Gaita Ari\~{n}o, Chris Greene, Levi Hall, Jennifer  Keller, Klaus Koepernick, and Erik  Ylvisaker for fruitful discussions.

\bibliographystyle{apsrev}
\bibliography{TmB4}

\end{document}